\documentclass[twoside,11pt]{article}
\usepackage{jair, rawfonts}

\usepackage{epsfig}
\usepackage{fancyhdr}
\usepackage{amsmath}
\DeclareMathOperator*{\argmin}{arg\,min}
\usepackage{amsfonts}
\usepackage{amssymb}
\usepackage{array}
\usepackage{graphicx}
\usepackage{url}
\usepackage{subfigure}
\usepackage{bm}
\usepackage{breqn}
\usepackage{xcolor}
\usepackage{soul}
\usepackage{amssymb}
\usepackage[breaklinks=true,hidelinks]{hyperref}
\usepackage{breakcites}

\usepackage{algorithm}
\usepackage{algpseudocode}
\usepackage{multirow}
\usepackage[justification=centering]{caption}
\usepackage{float}

\usepackage{graphics}

\newcolumntype{P}[1]{>{\centering\arraybackslash}p{#1}}
\newcolumntype{M}[1]{>{\centering\arraybackslash}m{#1}}

\graphicspath{{../figures}}

\jairheading
\ShortHeadings{SECRETS: Subject-Efficient Clinical Randomized Controlled Trials using Synthetic Intervention}
{Lala \& Jha}

\begin{document}

\date{}
\title{SECRETS: Subject-Efficient Clinical Randomized Controlled Trials using Synthetic Intervention}

\author{\name Sayeri Lala \email slala@princeton.edu \\
		\addr Department of Electrical and Computer Engineering, Princeton University \\
		Princeton, NJ 08544 USA \\
       \name Niraj K. Jha \email jha@princeton.edu \\
       \addr Department of Electrical and Computer Engineering, Princeton University \\
       Princeton, NJ 08544 USA  
       }


\maketitle

\begin{abstract}

The randomized controlled trial (RCT) is the gold standard for estimating the average treatment effect (ATE) of a medical intervention but requires 100s-1000s of subjects, making it expensive and difficult to implement. While a cross-over trial can reduce sample size requirements by measuring the treatment effect per individual, it is only applicable to chronic conditions and interventions whose effects dissipate rapidly. Another approach is to replace or augment data collected from an RCT with external data from prospective studies or prior RCTs, but it is vulnerable to confounders in the external or augmented data. We propose to simulate the cross-over trial to overcome its practical limitations while exploiting its strengths. We propose a novel framework, SECRETS, which, for the first time, estimates the individual treatment effect (ITE) per patient in the RCT study without using any external data by leveraging a state-of-the-art counterfactual estimation algorithm, called synthetic intervention. It also uses a new hypothesis testing strategy to determine whether the treatment has a clinically significant ATE based on the estimated ITEs. We show that SECRETS can improve the power of an RCT while maintaining comparable significance levels; in particular, on three real-world clinical RCTs (Phase-3 trials), SECRETS increases power over the baseline method by $\boldsymbol{6}$-$\boldsymbol{54\%}$ (average: 21.5\%, standard deviation: 15.8\%). 
\end{abstract}

\providecommand{\keywords}[1]
{
  \small	
  \textbf{\textit{Keywords---}} #1
}

\keywords{Clinical Randomized Controlled Trials, Counterfactual Estimation, Hypothesis Testing, Sample Efficiency, Synthetic Intervention.}

{
  \renewcommand{\thefootnote}{}%
  \footnotetext[1] {This work has been submitted to the IEEE for possible publication.}
  \footnotetext[2]{\copyright \space 20xx IEEE. Personal use of this material is permitted. Permission from IEEE must be obtained for all other uses, in any current or future
media, including reprinting/republishing this material for advertising or
promotional purposes, creating new collective works, for resale or
redistribution to servers or lists, or reuse of any copyrighted
component of this work in other works.}
}

\section{Introduction}

The randomized controlled trial (RCT) is the gold-standard approach to estimating the population-level or average 
treatment effect (ATE) of a drug, therapy, or other medical interventions 
\cite{friedman2015fundamentals, yao2021survey}. It is generally used in Phase-3 clinical studies 
to establish a treatment's effect \cite{friedman2015fundamentals,stanley2007design}. The typical RCT, i.e., the two-arm, parallel group, superiority 
trial \cite{friedman2015fundamentals}, evaluates whether a treatment is 
superior to standard treatment or placebo by estimating the ATE and determining whether the estimation 
is statistically significant. To estimate the ATE, the RCT first recruits subjects representing the population of 
interest and randomly divides them between a control arm and a treatment arm, in which subjects receive the standard 
care/placebo and treatment of interest, respectively. Then, it monitors a pre-specified health metric across both 
groups in parallel till the end of the study, evaluates the average treatment outcome (defined by the health metric) 
under each arm, and reports their difference as the ATE. The random allocation is crucial to accurately estimating 
the ATE because it removes confounders by producing more comparable cohorts between the control and treatment arms 
and allows for standard statistical hypothesis testing, i.e., two-sided test, to be used to detect target ATEs at desired 
accuracies \cite{friedman2015fundamentals}. However, as a consequence of between-subject variability in responses, 
RCTs typically require 100s-1000s of subjects to detect a clinically significant ATE with high accuracy 
\cite{stanley2007design,phased_trial}, making them expensive (at least \$42K per participant in Year 2021 on an average 
\cite{prasad2021reliable}) and difficult to implement, with recruitment generally being the ``most difficult 
task" \cite{friedman2015fundamentals}. 

One approach to making RCTs more subject-efficient is to use the cross-over design, which increases power over 
a parallel RCT by reducing the variance of the ATE estimate \cite{senn2002cross}. It does this by measuring the 
individual treatment effect (ITE), defined as the difference in the individual's response under the treatment and 
control conditions, per subject, and averaging over the ITEs. The variance of the ATE estimate is lower than that 
under the parallel RCT because the number of observations is doubled (two per subject) and the variance of the outcome 
measure is expectedly lowered by removing between-patient variability \cite{friedman2015fundamentals, senn2002cross}. 
While capable of lowering the sampling complexity relative to that of the parallel RCT 
\cite{senn2002cross,blackston2019comparison}, the cross-over design faces some constraints that 
preclude its widespread usage. To appreciate these limitations, we consider the simplest version, i.e., the two-period 
cross-over design. This design compares the two treatments (e.g., standard care/placebo and treatment of interest) 
by assigning each patient to a random sequence of treatments, in which the patient receives the standard care or 
treatment of interest in the first period and the opposite in the second period; the randomized sequence is used to 
rule out any temporal effects. Importantly, a washout phase is inserted between the two periods to remove any carryover 
effects from the treatment administered in the first period. The need for a washout restricts 
the types of condition-intervention pairs that can be studied under the cross-over, making it only suitable for 
chronic conditions for which treatments have effects that rapidly wash out \cite{friedman2015fundamentals,senn2002cross}.    

Instead of trying to estimate ATEs from small randomized samples, another approach tries to reduce the variance of the 
ATE estimate by using large external datasets obtained from previous clinical trials, electronic health records, patient registries, etc. \cite{thorlund2020synthetic} to augment or replace existing RCT data, since they can be cheaper and 
easier to acquire \cite{friedman2015fundamentals,prasad2021reliable} and more representative of the population compared to RCT data \cite{bica2021real}. However, ATE estimation 
from the integrated dataset is difficult because the data are no longer randomized. In particular, since external 
data can differ from the RCT data of interest in several ways, e.g., subject characteristics, medical protocol, etc., 
merging the two datasets can introduce confounders \cite{friedman2015fundamentals,thorlund2020synthetic}, i.e., variables that affect both 
the treatment response and treatment assignment, which produce uncomparable control and treatment groups and thereby 
preclude accurate ATE estimation \cite{yao2021survey}. Consequently, external control data need to be carefully curated, 
preferably from historical RCTs with comparable data collection methods, study endpoints, and study populations, to reduce discrepancies from the RCT 
of interest \cite{thorlund2020synthetic}. To learn in the presence of confounders, methods have been developed to 
estimate ATEs from observational data, including re-weighting to construct new comparable control and treatment 
groups \cite{thorlund2020synthetic} and stratification to condition on confounders when calculating the 
ATE \cite{yao2021survey}. In addition, counterfactual estimation algorithms have been developed to estimate 
ITEs \cite{yao2021survey}; however, these algorithms estimate the ITEs for the treatment group only, most likely 
because large, historical control data are available; consequently the estimated ATE is based only on estimated ITEs 
from the treatment group \cite{hojbjerre2022increasing,lim2018minimizing}. However, all these approaches to learning 
from observational data are only effective when there are no hidden confounders and when there is an overlap in the 
support (covariate distribution) between the control and treatment groups, conditions that are unlikely to hold since 
verifying that there are no hidden confounders would require domain expertise and operating over high-dimensional 
datasets reduces overlap \cite{bica2021real}. As a consequence of such conditions, attempts to estimate ATEs from 
observational data, whether combined with an existing RCT data or alone, require even larger sample sizes than 
RCTs \cite{prasad2021reliable,hojbjerre2022increasing,qian2021synctwin}.  

Instead of using external control datasets and facing the curation and analytical challenges, we propose to
simulate the cross-over design to leverage the variance reduction principle without facing its practical
constraints. Given an RCT dataset, our framework, dubbed SECRETS (\textbf{S}ubject-\textbf{E}fficient
\textbf{C}linical \textbf{R}andomiz\textbf{E}d Controlled \textbf{T}rials using \textbf{S}ynthetic Intervention), 
first estimates ITEs across \emph{all} subjects \emph{only} from the \emph{concurrent} control and treatment groups 
by using a state-of-the-art counterfactual estimation algorithm, i.e., synthetic intervention (SI) 
\cite{agarwal2023synthetic}, which is capable of estimating counterfactuals under control \emph{and} treatment conditions; 
then it averages the ITEs to estimate the ATE and applies a novel hypothesis testing algorithm to control the 
type-1 error rate
in order to detect a target ATE with high accuracy.

Our work differs from related work in several ways. While some studies have done ITE estimation without external 
control data, they use the estimated ITEs to identify subgroups within the RCT study group showing enhanced treatment 
effects \cite{foster2011subgroup,wolf2022permutation}, to support the 
primary findings of an RCT \cite{friedman2015fundamentals}. 
Instead, we use ITE estimation to improve power at a given sample size, thereby reduce the sample size needed to 
answer the primary question of the RCT, i.e., is the target ATE present in the study population; hence, the problem 
addressed in these prior works is orthogonal to the one we address. In addition, these works use the virtual twins 
(VT) algorithm \cite{foster2011subgroup} to estimate the counterfactuals used to estimate the ITEs. The VT algorithm 
fits regression models (e.g., random forests) on covariates to predict the counterfactual response of the patient 
under a treatment different from the one the patient receives. In contrast, our framework uses SI, a nonparametric 
algorithm that estimates the patient's counterfactual response under a given intervention using the observed responses across all patients assigned to that intervention.  

Since SI is the most relevant work, we describe how our study differs from it. First, while SI has been evaluated 
for enabling data-efficient RCTs \cite{agarwal2023synthetic}, it was used to reduce the number of \emph{interventions} 
performed per unit. Since standard clinical RCTs assign each subject to only one intervention, we instead show how SI can be used to reduce the number of subjects assigned to each intervention. Furthermore, the definition of the causal treatment effect and the 
metric used to assess data efficiency in the SI framework differ from those used in clinical RCTs. SI defines the 
causal treatment effect as the time-averaged response under a given intervention, while clinical RCTs define it to 
be the ATE or the difference in responses under the treatment and control interventions. While SI measures the 
squared error metric of the estimated causal parameter, we measure the significance level and power obtained to 
detect a target ATE, given that clinical RCTs set the sample size according to these values. 
Finally, the e-commerce RCT study used to evaluate SI does not resemble the data characteristics of clinical RCTs, 
potentially affecting the generalizability of their findings; in particular, in the e-commerce study, the 
pre-intervention data spanned eight timepoints, likely boosting performance \cite{amjad2018robust}, and the 
units were customer groups whose responses were averaged across individuals, potentially yielding less noisy donors. 
In contrast, clinical RCTs typically have one pre-intervention or baseline measurement and the units are individuals 
under our framework. 

Given the differences between our and prior/related work, we summarize our contributions as follows:

\begin{enumerate}
    \item We propose a novel framework, SECRETS, that, for the first time, improves power, hence 
subject efficiency, over the standard RCT \emph{without relying on any external data}. Specifically, first it 
estimates ITEs across \emph{all patients} within the concurrent control and treatment arms by using a state-of-the-art 
counterfactual estimation algorithm, SI, capable of counterfactual estimation across control and treatment conditions. 
Then it uses the estimated ITEs to reduce the variance of the ATE estimate, and implements a novel hypothesis testing 
procedure to control the type-1 error rate, given that the ITEs estimated under SI violate assumptions under standard 
hypothesis testing strategies.  
    
    \item We validate SECRETS on three real-world clinical RCT datasets, showing that it can boost power over the 
baseline approach by 6-54\% (average: 21.5\%, standard deviation: 15.8\%), thereby reduce the sample size needed to 
match a typical target statistical operating point (i.e., 5\% significance level, 80\% power) by 25-76\% or 10-3,957 
for both the control and treatment groups.

    \item We conduct ablation studies to illustrate the importance of the counterfactual estimation algorithm and 
the hypothesis testing strategy to the success of SECRETS.
    
\end{enumerate}
  
The rest of the article is organized as follows. We provide background on topics relevant to our proposed 
framework in Section \ref{sec:background}. We describe our proposed framework in Section \ref{sec:method}. 
We explain how we evaluate performance in Section \ref{sec:eval}. We present our results in Section \ref{sec:results}, 
discuss their implications in Section \ref{sec:discuss}, and draw conclusions in Section \ref{sec:con}.

\section{Background} \label{sec:background}

In this section, we review the concepts relevant to understanding SECRETS, i.e., the counterfactual estimation 
algorithm and hypothesis testing schemes.

\subsection{Synthetic Intervention for Counterfactual Outcome Estimation}

SI is a state-of-the-art counterfactual estimation algorithm that, for a given unit (e.g., patient), estimates 
counterfactual outcome trajectories under interventions different from the one that the unit was exposed to during 
the intervention period \cite{agarwal2023synthetic}. To do this, first, it assumes that a pool of donor units exposed to 
the interventions (including the control) exists and that each unit has been assigned to the control arm during the 
pre-intervention period. Then, for a given target unit observed under some intervention $j$, SI calculates its 
counterfactual outcome under a different intervention $k$ by weighting the post-intervention outcomes across the donor units exposed to intervention $k$. It uses principal component regression to learn weights 
over the donor units such that the weighted sum of their pre-intervention outcomes best predicts the target unit's 
pre-intervention outcome. SI's efficacy over other methods, like matching and re-weighting \cite{yao2021survey}, stems from its ability to 
predict counterfactuals under the presence of unobserved confounders \cite{agarwal2023synthetic}. 

\subsection{Hypothesis Testing}

Hypothesis testing is a statistical procedure used to determine whether a parameter of interest $\mu$, e.g., the 
ATE of a drug, is some specified, non-zero value. The standard setup for a continuous parameter considers two 
hypotheses: null, which posits that $\mu=\mu_{0}=0$, and the alternative, which posits that $\mu=\mu_{1} > 0$ (under 
the two-sided test where we are interested in the magnitude) \cite{friedman2015fundamentals}. The objective is to design a test that maximizes the 
power $1-\beta$ (the probability of identifying that $\mu=\mu_{1}$ when the alternative hypothesis is true) while 
capping the type-1 error or significance level $\alpha$ (the probability of identifying that $\mu=\mu_{1}$ when the 
null hypothesis is true). The test is defined by a test statistic calculated from the observations and a rejection region determined by critical values set to control the significance level; it rejects the null hypothesis if the statistic falls in the rejection region and otherwise 
fails to reject the null hypothesis.

Test strategies differ in how they choose the test statistic and rejection region 
in order to improve power, based on assumptions about the distribution of observations under the null hypothesis. For example, the two-sided \emph{t}-test for the mean of a normal distribution assumes that observations are 
independently and identically distributed according to a normal distribution with unknown variance and sets the test statistic to be the sample mean normalized by the sample variance and the critical value to be the $(1-\alpha/2)$th percentile of a Student's \emph{t} distribution (with appropriate degrees of freedom), where $\alpha$ is the target significance level. When the 
data distribution is unknown, the bootstrap test resamples from the data to set the critical value(s) according to $\alpha$ \cite{rosner2015fundamentals,good_2005}. 

However, these testing strategies assume that 
observations in the data sample are independent. When independence fails to hold, applying these tests results in 
higher significance levels than desired. To address this, several modifications have been proposed 
\cite{dale2002spatial}. One approach attempts to correct the estimate of sample variance used in the statistic by 
accounting for the covariance terms induced by the data dependencies; however, this strategy was empirically shown 
to yield high variance estimates when the correlational structure was estimated from the data 
\cite{dale2002spatial}. Another approach tries to improve estimates of the correlational structure 
by fitting autoregressive models to model the data dependencies and sampling from them to construct confidence 
intervals for testing; however, this approach may be inefficient when operating over large sample 
sizes \cite{dale2002spatial}. Instead of directly estimating the correlational structure, restricted randomization 
methods construct a null distribution that accounts for the dependency structure; however, these methods assume 
special dependency structures in the data \cite{fortin2000randomization}.

\section{Methodology} \label{sec:method}

Having reviewed the necessary background, we now present our framework, SECRETS. First, we provide an overview of the 
framework and explain how it reduces the sampling complexity of a standard RCT, i.e., the two-arm, parallel group, superiority 
trial. Then, we describe each step of the framework in detail.

\subsection*{Overview}

\begin{figure}[hbt!]
    \centering
    \includegraphics[scale=0.5]{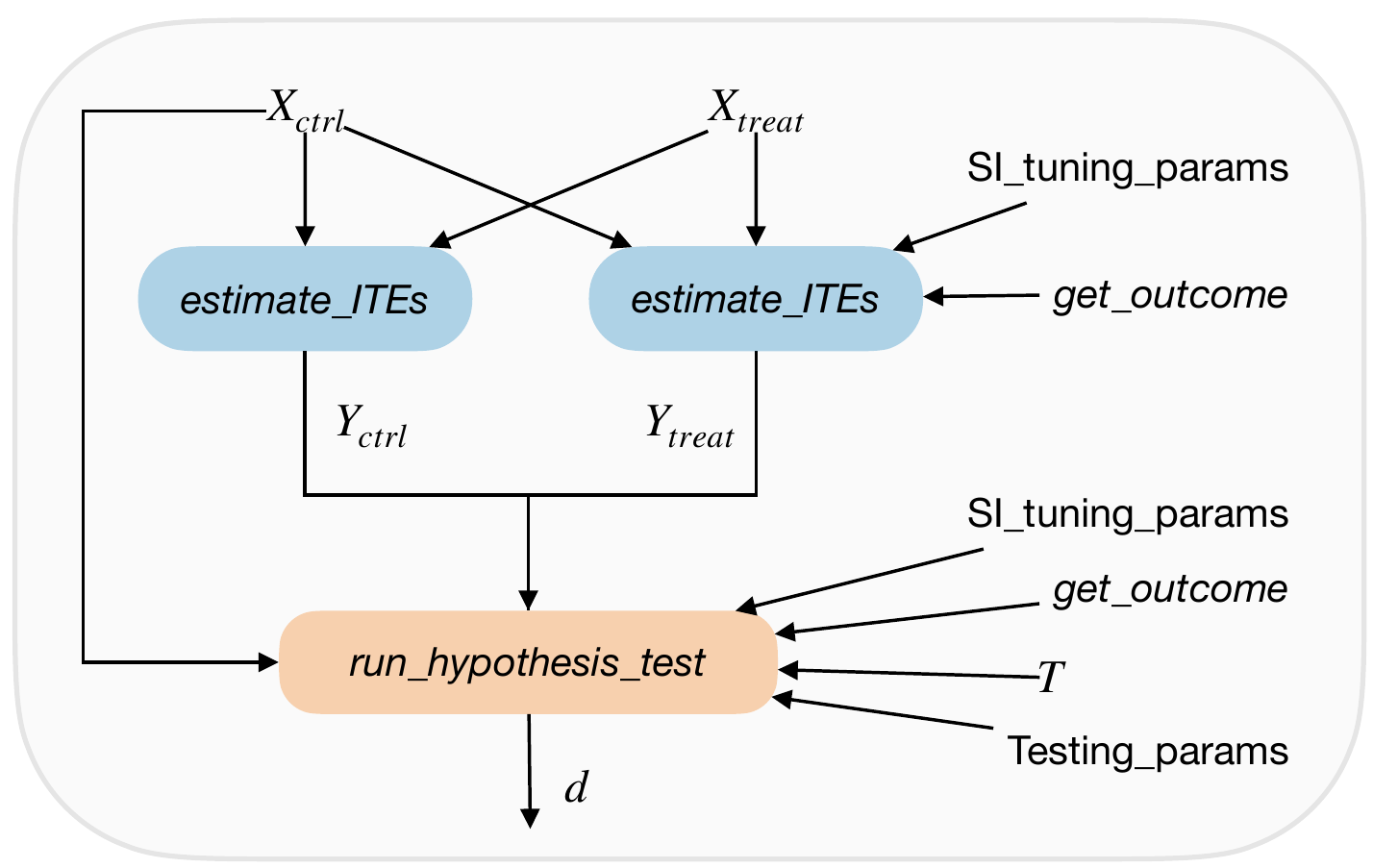}
    \caption{Flowchart of the SECRETS framework. Note that both calls to \emph{estimate\_ITEs} take in
SI\_tuning\_params and \emph{get\_outcome} but we have omitted the arguments for brevity.}
    \label{fig:secrets}
\end{figure}

The SECRETS framework, illustrated in Fig.~\ref{fig:secrets}, 
improves power at a given
sample size and thereby reduces the sample size required to reach a target power $1-\beta_{target}$ for a target ATE $\mu_{1,target}$ and
significance level $\alpha_{target}$, compared to the standard RCT.
It simulates the cross-over design in order to reduce the variance of the ATE estimate. Given
data collected from an already-conducted RCT, i.e., the observed control data $X_{ctrl}$ and treatment
data $X_{treat}$, parameters for SI tuning, and a function \emph{get\_outcome} that calculates the health outcome of interest from a patient's response trajectory under some intervention, SECRETS first estimates the ITEs across all patients per arm, i.e., $Y_{ctrl}$ and $Y_{treat}$. Then it implements and runs a two-sided hypothesis test using the control data, the merged ITEs, the parameters for SI tuning, the function \emph{get\_outcome}, the number of samples to generate from the null distribution ($T$), and additional parameters for testing (see Appendix \ref{sec:supp-alg}, Alg. \ref{alg:main} for input descriptions). 
Next, we describe each step of SECRETS in detail.   

\subsection*{Step 1: Estimate the Individual Treatment Effects}

\begin{figure*}[hbt!]
    \centering
    \includegraphics[scale=0.5]{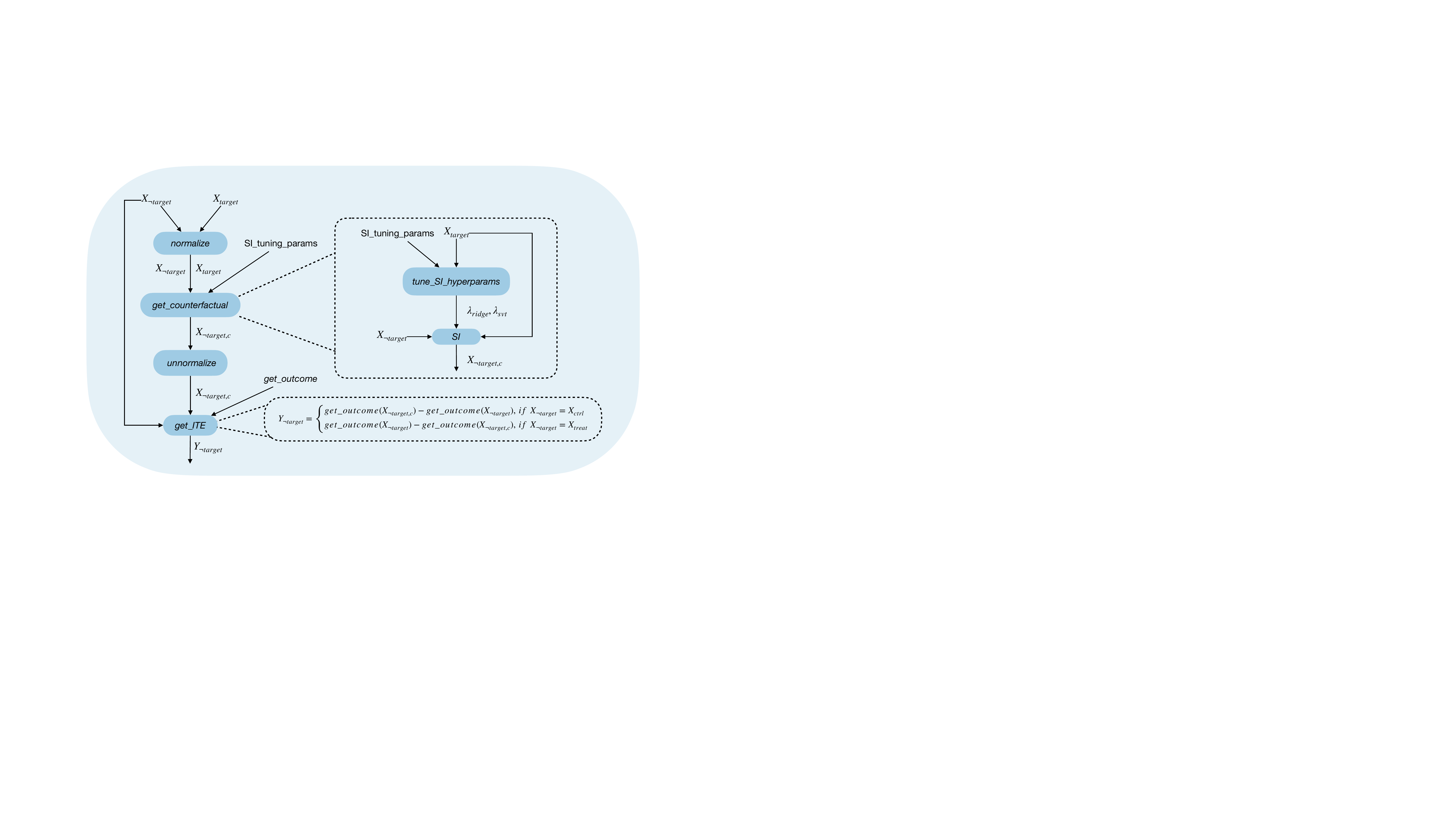}
    \caption{Flowchart of the subroutine, \emph{estimate\_ITEs}.}
    \label{fig:est_ites}
\end{figure*}

To estimate the ITEs $Y_{ctrl}$ and $Y_{treat}$, SECRETS uses the subroutine \emph{estimate\_ITEs}, shown in Fig. \ref{fig:est_ites}. Given data from an arm exposed to the target intervention ($X_{target}$) and data from an ``unexposed" arm ($X_{\neg target}$), i.e., one exposed to a different intervention, \emph{estimate\_ITEs} first performs min-max normalization and calls \emph{get\_counterfactual} to impute the response of the unexposed arm under the target intervention ($X_{\neg target,c}$). To do this, \emph{get\_counterfactual} uses SI by first tuning its hyperparameters ($\lambda_{ridge}$ and $\lambda_{svt}$) on $X_{target}$ with the subroutine \emph{tune\_SI\_hyperparams} and the SI tuning parameters (see Appendix \ref{sec:supp-alg}, Alg. \ref{alg:tune_si_hparams}) and then running \emph{SI} with $X_{target}$ as the donor group, the tuned hyperparameters, and each unit from $X_{\neg target}$ as the target unit (see Appendix \ref{sec:supp-alg}, Alg. \ref{alg:si}). After unnormalizing $X_{\neg target,c}$, \emph{estimate\_ITEs} calls \emph{get\_ITE}, which takes the difference between the outcomes under the target and observed interventions, calculated with \emph{get\_outcome} (vectorized) on $X_{\neg target,c}$ and $X_{\neg target}$, respectively. In particular, SECRETS calculates $Y_{ctrl}$ by setting $X_{\neg target}$ to $X_{ctrl}$ and $X_{target}$ to $X_{treat}$ and vice-versa to calculate $Y_{treat}$.

\subsection*{Step 2: Conduct a Data-driven Hypothesis Test}

\begin{figure*}[hbt!]
    \centering
    \includegraphics[scale=0.5]{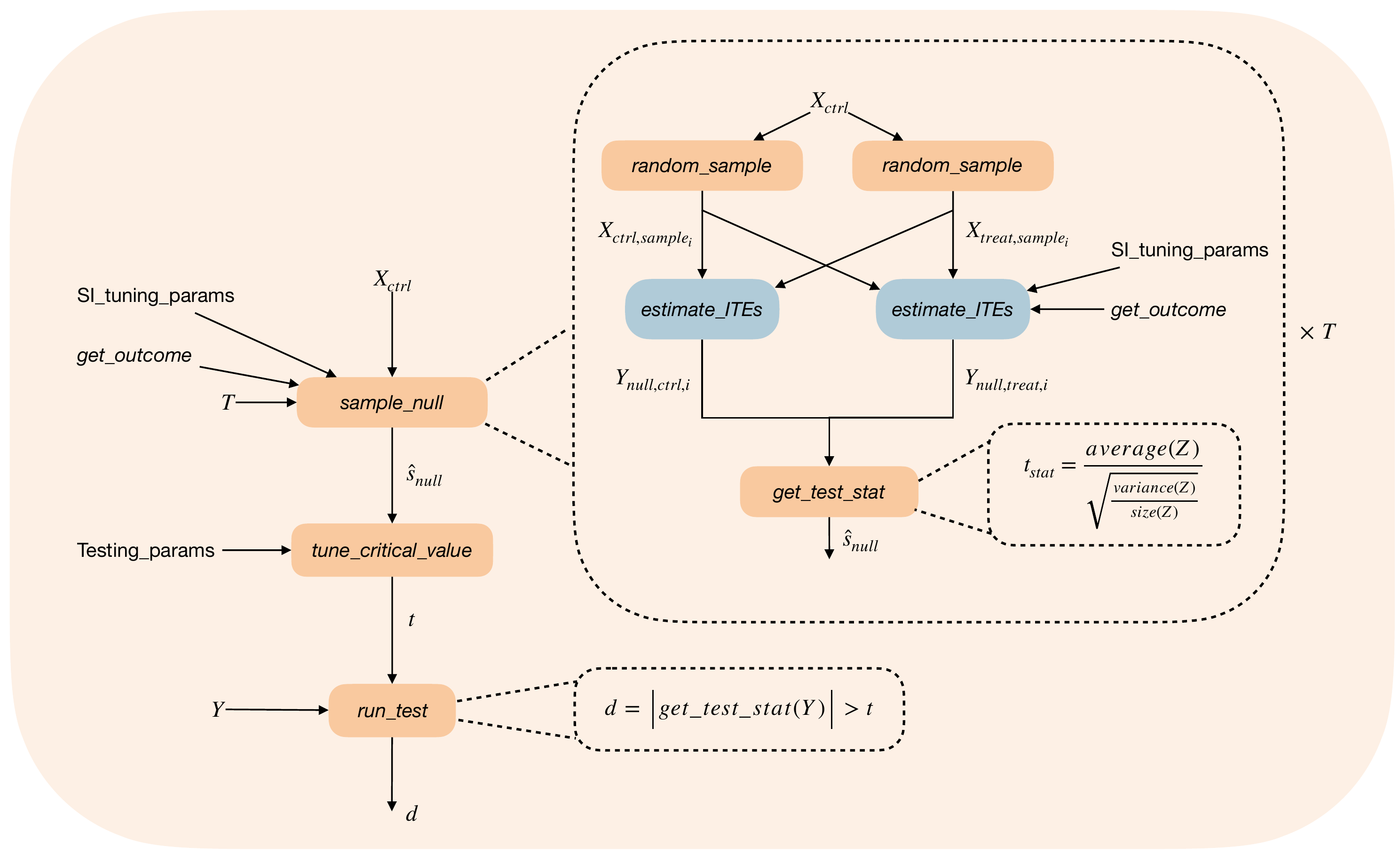}
    \caption{Flowchart of the subroutine, \emph{run\_hypothesis\_test}. Note that both calls to
\emph{estimate\_ITEs} in \emph{sample\_null} take in the SI\_tuning\_params and \emph{get\_outcome} function
but we have omitted the arguments for brevity.}
    \label{fig:run_hyp}
\end{figure*}

After calculating the ITEs, SECRETS conducts a hypothesis test that uses the ITEs to determine whether to reject or not reject the 
null hypothesis. The standard two-sided one-sample tests 
are unsuitable because the estimated ITEs are not independent of each other, given that SI uses the same donor pool, i.e., the treatment (control) group, to estimate the counterfactuals per target unit in the control (treatment) group. In addition, existing 
methods to deal with dependent samples either perform poorly in practice or make assumptions about the dependency 
structure that are not applicable to the estimated ITEs.  
Therefore, SECRETS uses a new hypothesis testing procedure \emph{run\_hypothesis\_test}, shown in Fig. \ref{fig:run_hyp}, that accounts for the dependencies of the ITEs without facing these limitations. 

In particular, \emph{run\_hypothesis\_test} implements the critical-value test procedure used in the two-sided one-sample 
\emph{t}-test but uses the data to tune the critical value $t$. To tune $t$, \emph{run\_hypothesis\_test} first 
approximates the null distribution of the test statistic ($\hat{s}_{null}$) by sampling from it $T$ times
with $sample\_null$. To generate the $i$-th sample, \emph{sample\_null} first samples with replacement from
the original control data ($X_{ctrl}$) using \emph{random\_sample} to construct control and treatment groups
with an equal number of subjects as the original control and treatment groups and with comparable responses
($X_{ctrl,sample_i}$ and $X_{treat,sample_i}$). Then, as in step 1, it runs \emph{estimate\_ITEs} (with the
SI tuning parameters and function \emph{get\_outcome}) to calculate the ITEs for the constructed control and
treatment groups. Since the constructed control and treatment groups were exposed to the same intervention
(i.e., the control condition), the corresponding ITEs $Y_{null,ctrl,i}$ and $Y_{null,treat,i}$ are samples
from the null distribution. Given the merged set of ITEs, \emph{get\_test\_stat} calculates the resulting
test statistic $\hat{s}_{null,i}$ using the test statistic formula from the one-sample \emph{t}-test. 

Afterwards, \emph{run\_hypothesis\_test} tunes the test's critical value with \emph{tune\_critical\_value},
which uses $\hat{s}_{null}$ and a set of testing parameters that includes the target significance level $\alpha_{target}$ and search range parameters to find the critical value $t$ attaining the target significance level, in a fashion similar to binary search (see Appendix \ref{sec:supp-alg}, Alg. \ref{alg:get_cv}). 

Finally, \emph{run\_hypothesis\_test} runs the two-sided test by calculating the test statistic from the ITEs derived from the original control and treatment groups ($Y$) in step 1 and comparing its magnitude against the tuned critical value $t$, rejecting the null hypothesis if $d=1$ and otherwise failing to reject it.    

\section{Performance Evaluation Setup} \label{sec:eval}

In this section, we describe the experimental setup used to evaluate SECRETS. First, we present the performance 
metrics. Then, we describe the baseline method and ablations used to understand the framework. Finally, we describe 
the datasets used to conduct the experiments.  

\subsection*{Metrics}

To assess sample efficiency, we compare the powers obtained for a given sample size of $n_{a}$ subjects per arm 
(i.e., control and treatment groups) and target significance level $\alpha_{target}$. We also report the sample size 
required to obtain a desired power $1-\beta_{target}$. To measure power $1-\beta$ and significance level $\alpha$, 
we follow the approach from \cite{blackston2019comparison}, which simulates many trials under the alternative and 
null settings and calculates the percentage of trials where the test procedure returns a reject, respectively. 
First, we define the target ATE under the alternative setting, i.e., $\mu=\mu_{1,target}$, as the ATE measured on 
the full RCT dataset, and define $\mu=\mu_{0}=0$ as the ATE under the null setting. Then, to measure power 
$1-\beta$ given a sample of size $n_{a}$ subjects per arm, we run $L$ trials; we simulate a trial under the 
alternative setting by constructing new control and treatment groups by sampling $n_{a}$ subjects with replacement 
from the original RCT's control and treatment groups, respectively. Likewise, we measure the significance level 
$\alpha$ with $L$ trials by simulating the null setting, in which we construct both the control and treatment groups 
by sampling $n_{a}$ subjects with replacement from the original RCT's control group. 

To understand how changes in the distribution (mean and variance) of the estimated ATE, i.e., the sample statistic, 
under the alternative and null settings contribute to changes in power, we consider the power equation for a 
two-sided hypothesis test \cite{rosner2015fundamentals}, i.e., Eq.~(\ref{eq:power}), as a model. 

\begin{equation} \label{eq:power}
    1-\hat{\beta}\approx\Phi\Biggr[{-z_{1-\alpha_{target}/2}+\frac{|\hat{\mu}_{0}-\hat{\mu}_{1}|}{\hat{\sigma}}}\Biggr]
\end{equation}

The equation states that if the underlying distribution of the sample statistic is normal, then the power of the 
two-sided \emph{z}-test, $1-\hat{\beta}$, is approximately given by the cumulative distribution function $\Phi$, 
evaluated at the negative of the critical value $z_{1-\alpha_{target}/2}$, i.e., the $(1-\alpha_{target}/2)\%$ of a standard normal 
distribution, shifted by a constant, which we refer to as the \emph{shift term}. The \emph{shift term} is given 
by Eq.~(\ref{eq:shift_factor}), where parameters $\hat{\mu}_{0}$ and $\hat{\mu}_{1}$ correspond to the means of 
the distribution of the sample statistic under the null and alternative settings, respectively, and parameter 
$\hat{\sigma}$ is the standard deviation of the distributions. The equation implies that for a given 
$\alpha_{target}$, power increases as \emph{shift term} increases, which occurs if the means get further separated 
and/or the variance of the distributions decreases. Hence, if the equation estimates power well, we can interpret 
differences in power in terms of differences in the means and variance of the distributions. 

\begin{equation} \label{eq:shift_factor}
    \text{\emph{shift term}}=\frac{|\hat{\mu}_{0}-\hat{\mu}_{1}|}{\hat{\sigma}}
\end{equation}

In our analysis, $\hat{\mu}_{0}$ and $\hat{\mu}_{1}$ correspond to the mean of the distribution of the ATE
estimate (derived from the trials) under the null and alternative settings, respectively. We estimate
$\hat{\sigma}$ by averaging over the standard deviations of the distribution of the ATE estimate under the null and
alternative settings and also report the standard deviation in this estimate (std.) to show that the
standard deviations of the distributions are comparable. To assess whether the model holds, we report the model's
estimate of power $1-\hat{\beta}$ given $\alpha_{target}$, $\hat{\mu}_{0}$, $\hat{\mu}_{1}$, and
$\hat{\sigma}$. While the equation may not predict measured power $1-\beta$ exactly because of 
non-normality of the ATE estimate under a small sample size and differences in the test procedure, we can still use it to compare methods 
and qualitatively assess how differences in distributions explain differences in power.  

We also assess the quality of the estimated ATEs since they may be of interest \cite{friedman2015fundamentals}. We 
calculate the errors in the ATE estimates under the null and alternative settings by measuring the difference between 
the means of the distribution of the estimated ATEs from their corresponding true values, given by 
$\hat{\mu}_{0}-\mu_{0}$ and $\hat{\mu}_{1}-\mu_{1,target}$, respectively.

\subsection*{Baseline}

\begin{figure}[hbt!]
    \centering
    \includegraphics[scale=0.5]{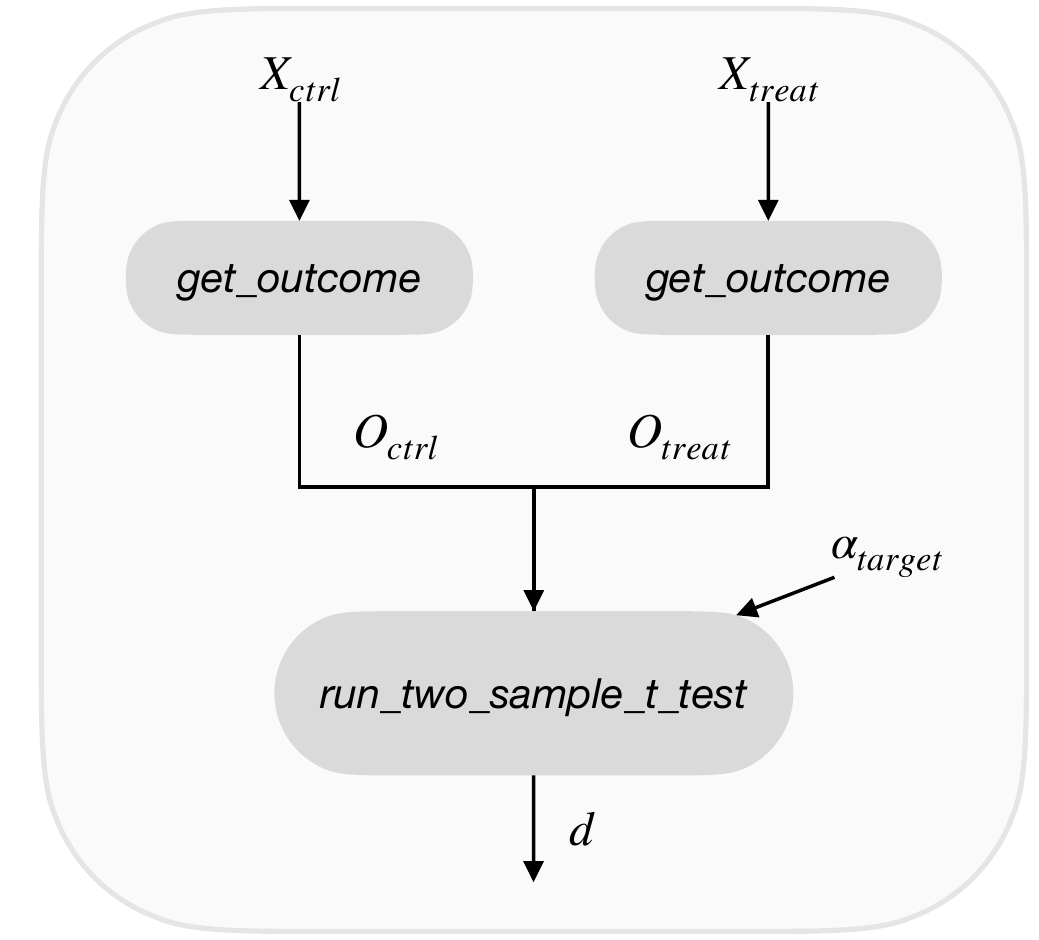}
    \caption{Flowchart of the \emph{Standard} baseline.}
    \label{fig:baseline}
\end{figure}

We describe the baseline method against which SECRETS is benchmarked. The baseline method, which we refer to as
\emph{Standard}, is the approach used by two-arm, parallel group, superiority RCT studies to determine a treatment's effect. Its flowchart is shown in Fig.~\ref{fig:baseline}. First, given the control and treatment data, it calculates the corresponding outcome data $O_{ctrl}$ and $O_{treat}$ using $get\_outcome$ (vectorized). Then it conducts the two-sample \emph{t}-test for
independent samples with unequal variances \cite{rosner2015fundamentals} using the outcome data and desired significance level $\alpha_{target}$.

\emph{Standard} estimates the sample size $n_{a}$ required per arm to 
obtain power $1-\beta_{target}$ for a target ATE of $\mu_{1,target}$ and significance level $\alpha_{target}$ using the equation based on the
two-sided \emph{z}-test \cite{rosner2015fundamentals}, i.e., Eq.~(\ref{eq:sample_size}), where $z_{m}$ refers to 
$m$-th percentile of the standard normal distribution, and $\sigma_{ctrl}^{2}$ and $\sigma_{treat}^{2}$ are the 
variances of the outcome measure for the control and treatment groups, respectively, typically estimated from 
prior studies \cite{friedman2015fundamentals}. 

\begin{equation}
    \label{eq:sample_size}
    n_{a}=\frac{(\sigma_{ctrl}^{2}+\sigma_{treat}^{2})(z_{1-\alpha_{target}/2}+z_{1-\beta_{target}})^{2}}{(\mu_{1,target}-\mu_{0})^{2}}
\end{equation}

For our experiments, we estimate $\sigma_{ctrl}^{2}$ and $\sigma_{treat}^{2}$ using the full dataset to obtain accurate sample size estimates. However, since the 
equation fails to hold for estimates based on small datasets \cite{friedman2015fundamentals, rosner2015fundamentals}, 
in such cases, we increase the sample size till measured power $1-\beta$ matches $1-\beta_{target}$.

\subsection*{Ablation Studies}

In this section, we describe the ablation studies used to evaluate the efficacy of each step underlying SECRETS. 

First, we run ablations on \textproc{Estimate\_Ites} by swapping the counterfactual estimation algorithm of
SECRETS, i.e., SI, with the VT algorithm \cite{foster2011subgroup}, which estimates counterfactual 
outcomes using a regression model fit on covariates. We conduct this ablation to evaluate the advantage of SI, which 
learns patient-specific models rather than shared models. In particular, we fit two shared models, i.e., one model 
to predict responses under the control condition and one to predict responses under the treatment condition, using 
all the patient data from control and treatment groups, respectively, and use the models to predict the 
post-intervention counterfactual response of each patient under the treatment and control groups, respectively. 
As with SI, we use the ridge regression estimator and tune the regularization strength using the validation set 
performance. We refer to this version of the framework as SECRETS-VT. 

Next, we run ablations on \textproc{Run\_Data\_Driven\_Hypothesis\_Test} by swapping the hypothesis testing algorithm 
of SECRETS with standard ones. We conduct this ablation to evaluate the need for a suitable hypothesis testing 
strategy, given the dependencies among the estimated ITEs. First, we try the 
one-sample \emph{t}-test for the mean of a normal distribution \cite{rosner2015fundamentals}, which we denote as 
SECRETS-T; this test assumes that the underlying samples (i.e., the estimate ITEs) are independent and identically distributed from a Gaussian 
distribution. Then, to assess if non-Gaussianity of the samples is an issue, we try the bootstrap hypothesis test, 
which we denote as SECRETS-B; this test still assumes that the samples are independent but relaxes the normality 
assumption, only requiring that samples be drawn from populations where the parameter being estimated is the same 
under the null hypothesis \cite{good_2005}. We specifically use the bias-corrected and accelerated version 
\cite{efron1987better} since it yields better estimates of the parameter of interest \cite{good_2005}. Then, to 
assess whether data dependence across samples is an issue, we try a variant of SECRETS-T, called SECRETS-T-P, in 
which we first permute samples across the trials to remove dependencies and then run the one-sample \emph{t}-test. 
After comparing against these standard test strategies, we assess how well the proposed hypothesis testing strategy 
performs by comparing against an ``oracle," which we denote as SECRETS-O. While SECRETS constructs the null 
distribution \emph{per} trial (via the subroutine \emph{sample\_null}), SECRETS-O constructs the null distribution using the 
estimated ITEs across \emph{all} the trials under the null setting, allowing it to identify the correct critical value 
that attains the significance level of $\alpha_{target}$. 

\subsection*{Clinical RCT Datasets}
We evaluate the 
framework on Phase-3 RCTs as they typically require 100s-1000s of subjects to establish a treatment's effect. We 
focus on the two-arm, parallel group, superiority trial design typically adopted in clinical RCTs 
\cite{friedman2015fundamentals} to assess the general utility of our method. We obtained datasets 
for three such RCTs, i.e., CHAMP (NCT01581281) \cite{powers2017trial,champ_stats}, ICARE (NCT00871715) \cite{winstein2016effect, icare_stats}, and MGTX (NCT00294658) \cite{wolfe2016randomized}, from the National Institute of Neurologic Disease and Stroke (NINDS) \cite{ninds_archive}, which are described in Appendix \ref{sec:supp-datasets}. 

 \subsection*{Implementation Details}

In this section, we specify the evaluation parameters and computing resources used to 
perform our experiments. Algorithmic parameters are detailed in Appendix \ref{sec:supp-alg_params}.

For evaluation, we set $\alpha_{target}$ to 5\%, and, for sample size requirements, we additionally set 
$1-\beta_{target}$ to 80\%, since this statistical operating point is commonly used in clinical RCTs 
\cite{friedman2015fundamentals}. We set the number of trials $L$ used to measure power $1-\beta$ and significance 
level $\alpha$ to 1000, since we empirically found this sufficient for the methods to converge close to $\alpha_{target}$ and $1-\beta_{target}$ within reasonable computation time. For the null 
setting, we set $\mu=\mu_{0}=0$, and for the alternative settings, we set $\mu=\mu_{1,target}$, with $\mu_{1,target}$ 
given by the ATE measured on each dataset, i.e., -3.17 for CHAMP, -3.00 for ICARE, and -2.70 for MGTX. 

To run the experiments, we used 28-32 CPU cores, 4GB of memory per CPU, and 2.4 GHz Intel Broadwell and 2.6 GHz 
Intel Skylake processors. We implemented the framework and experiments with Python using standard numerical packages.

\section{Experimental Results} \label{sec:results}

This section presents the results from our experiments. First we compare SECRETS against the baseline method 
\emph{Standard}. Then, to understand the importance of each component of SECRETS, we evaluate results from the 
ablation studies.

\subsection*{SECRETS vs. Baseline}

\begin{table*}[h!]
\centering
\caption{\emph{Standard} vs. SECRETS and SECRETS vs. SECRETS-VT on the CHAMP Dataset}
\label{tab:champ} 
\resizebox{\columnwidth}{!}{%
\begin{tabular}{|M{0.05\linewidth} M{0.17\linewidth} M{0.12\linewidth} M{0.075\linewidth} M{0.12\linewidth} M{0.1\linewidth} M{0.15\linewidth} M{0.1\linewidth} M{0.14\linewidth} M{0.1\linewidth}|}
\hline

${n_{a}}$ & {Method} & ${1-\beta}$ (\%) & ${\alpha}$ (\%) & ${1-\hat{\beta}}$ (\%) &
${|\hat{\mu}_{1}-\hat{\mu}_{0}|}$ & $\hat{\mu}_{1}-\mu_{1,target}$ & $\hat{\mu}_{0}-\mu_{0}$ & ${\hat{\sigma}}$ (std.) & {\emph{shift term}}  \\
\hline

\multirow[c]{3}{*}{550} & \emph{Standard} & 53 & 5.0 & 52 & 3.27 & -0.06 & 0.03 & 1.63 (0.01) & 2.01   \\ 
& SECRETS & 78 & 5.1 & 73 & 1.85 & 1.31 & -0.01 & 0.72 (0.04) & 2.58  \\
& SECRETS-VT & 58 & 5.4 & 57 & 3.08 & 0.07 & -0.02 & 1.43 (0.06) & 2.15 \\ \hline 

\multirow[c]{3}{*}{1130} & \emph{Standard} & 81 & 3.8 & 84 & 3.20 & -0.06 & -0.03 & 1.09 (0.04) & 2.93 \\
& SECRETS & 97 & 6.3 & 95 & 2.03 & 1.15 & 0.01 & 0.56 (0.04) & 3.62 \\
& SECRETS-VT & 91 & 5.8 & 90 & 3.10 & 0.05 & -0.02 & 0.95 (0.03) & 3.25 \\ \hline 

\end{tabular}%
}
\end{table*}

\begin{table*}[h!]
\centering
\caption{\emph{Standard} vs. SECRETS and SECRETS vs. SECRETS-VT on the ICARE Dataset}
\label{tab:icare} 
\resizebox{\columnwidth}{!}{%
\begin{tabular}{|M{0.05\linewidth} M{0.17\linewidth} M{0.12\linewidth} M{0.075\linewidth} M{0.12\linewidth} M{0.1\linewidth} M{0.15\linewidth} M{0.1\linewidth} M{0.14\linewidth} M{0.1\linewidth}|}
\hline

${n_{a}}$ & {Method} & ${1-\beta}$ (\%) & ${\alpha}$ (\%) & ${1-\hat{\beta}}$ (\%) &
${|\hat{\mu}_{1}-\hat{\mu}_{0}|}$ & $\hat{\mu}_{1}-\mu_{1,target}$ & $\hat{\mu}_{0}-\mu_{0}$ & ${\hat{\sigma}}$ (std.) & {\emph{shift term}}  \\
\hline

\multirow[c]{3}{*}{1250} & \emph{Standard} & 27 & 4.1 & 30 & 3.02 & -0.01 & 0.01 & 2.12 (0.03) & 1.43 \\
& SECRETS & 81 & 5.5 & 81 & 3.37 & -0.33 & 0.04 & 1.19 (0.03) & 2.83 \\
& SECRETS-VT & 73 & 6.5 & 72 & 4.96 & -1.93 & 0.04 & 1.94 (0.02) & 2.56 \\  
 \hline 

\multirow[c]{3}{*}{5207} & \emph{Standard} & 81 & 5.8 & 81 & 2.98 & -0.04 & -0.06 & 1.05 (0.01) & 2.83 \\
& SECRETS & 100 & 5.7 & 100 & 3.74 & -0.77 & -0.02 & 0.67 (0.07) & 5.56 \\
& SECRETS-VT & 100 & 6.2 & 100 & 4.87 & -1.91 & -0.04 & 0.94 (0.01) & 5.16 \\ \hline 

\end{tabular}%
}
\end{table*}

\begin{table*}[h]
\centering
\caption{\emph{Standard} vs. SECRETS and SECRETS vs. SECRETS-VT on the MGTX Dataset}
\label{tab:mgtx} 
\resizebox{\columnwidth}{!}{%
\begin{tabular}{|M{0.05\linewidth} M{0.17\linewidth} M{0.12\linewidth} M{0.075\linewidth} M{0.12\linewidth} M{0.1\linewidth} M{0.15\linewidth} M{0.1\linewidth} M{0.14\linewidth} M{0.1\linewidth}|}
\hline

${n_{a}}$ & {Method} & ${1-\beta}$ (\%) & ${\alpha}$ (\%) & ${1-\hat{\beta}}$ (\%) &
${|\hat{\mu}_{1}-\hat{\mu}_{0}|}$ & $\hat{\mu}_{1}-\mu_{1,target}$ & $\hat{\mu}_{0}-\mu_{0}$ & ${\hat{\sigma}}$ (std.) & {\emph{shift term}}  \\ \hline

\multirow[c]{3}{*}{30} & \emph{Standard} & 70 & 5.1 & 63 & 2.69 & -0.09 & -0.10 & 1.17 (0.06) & 2.30  \\
& SECRETS & 79 & 5.7 & 69 & 1.33 &  1.33 & -0.04 & 0.54 (0.03) & 2.45  \\
& SECRETS-VT & 71 & 5.1 & 77 & 2.49 & 0.17 & -0.04 & 0.92 (0.01) & 2.70 \\  \hline  

\multirow[c]{3}{*}{40} & \emph{Standard} & 81 & 4.8 & 75 & 2.67 & -0.06 & -0.09 & 1.01 (0.02) & 2.63  \\ 
& SECRETS & 87 & 5.2 & 80 & 1.32 & 1.34 & -0.04 & 0.47 (0.01) & 2.80 \\
& SECRETS-VT & 85 & 5.5 & 87 & 2.45 & 0.20 & -0.04 & 0.79 (0.02) & 3.09 \\  \hline 

\end{tabular}%
}
\end{table*}

Across all datasets, SECRETS obtains better power and comparable significance level compared to \emph{Standard}, and the 
gains in power translate into reductions in the number of subjects required to obtain a desired power 
$1-\beta_{target}$ and significance level $\alpha_{target}$. SECRETS is able to increase power by consistently reducing the 
variance of the ATE estimate, 
albeit at the cost of higher error in the ATE estimate 
under the alternative setting. For example, on the CHAMP dataset (Table \ref{tab:champ}), for an arm size ($n_{a}$) of 
550, \emph{Standard} has a power ($1-\beta$) of 53\% and a significance level ($\alpha$) of 5.0\% while SECRETS has a 
power of 78\% and a significance level of 5.1\%. 
By increasing the power at a given arm size, SECRETS is able to converge close to the desired statistical operating
point of 80\% power ($1-\beta_{target})$ and 5\% significance level ($\alpha_{target}$) at an arm size of 550
while \emph{Standard} does so at an arm size of 1130. Hence, SECRETS reduces the number of required samples by
almost 51\% or 580 subjects per arm. 

To assess the factors contributing to the increased power under SECRETS, we
analyze the distributions of the ATE estimates using the power estimation model (Eq.~(\ref{eq:power})), given that it estimates the measured power well, e.g., under an arm size 
of 550, the model estimates a power $(1-\hat{\beta})$ of 52\% and 73\% compared to the measured powers $(1-\beta)$ of 
53\% and 78\% for \emph{Standard} and SECRETS, respectively. First, we note that the means are further separated under \emph{Standard}, i.e., the 
distance between the means ($|\hat{\mu}_{1}-\hat{\mu}_{0}|$) is 3.27 and 1.85 under \emph{Standard} and SECRETS, 
respectively. Compared to SECRETS, \emph{Standard} better separates the means because it has lower error in the ATE 
estimate under the alternative setting ($\hat{\mu}_{1}-\mu_{1,target}$) while both have comparably low error in the 
ATE estimate under the null setting ($\hat{\mu}_{0}-\mu_{0}$). 
 Despite their enhanced mean separation under 
\emph{Standard}, the distributions have lower variance under SECRETS than under \emph{Standard}, i.e., the average 
standard deviation ($\hat{\sigma}$) is 1.63 and 0.72 under \emph{Standard} and SECRETS, respectively. The lower variance 
under SECRETS outweighs the advantage of better mean separation under \emph{Standard}, giving SECRETS a higher 
\emph{shift term} of 2.58 compared to 2.01 under \emph{Standard}, which translates into higher power. Similar trends hold for an arm size of 1130, as well as on the ICARE (Table \ref{tab:icare}) and MGTX datasets (Table \ref{tab:mgtx}). 

On some datasets, \emph{Standard} appears to have significance levels below the target 5\%  while SECRETS appears to have significance levels slightly above it (e.g., CHAMP with $n_{a}$ of 1130 in Table \ref{tab:champ}), a performance gap which might explain why SECRETS has higher power than \emph{Standard}. However, this gap in the significance levels is explained away by experimental limitations and is expected to vanish upon removing them while maintaining the power gains under SECRETS. In particular, increasing the number of trials $L$ enables \emph{Standard} to converge to the 5\% significance level while maintaining comparable power (see Appendix \ref{sec:supp-tables}, Table \ref{tab:num_trials}). In addition, increasing $T$ or the number of samples generated under the null distribution under SECRETS enables it to converge to 5\% significance level while maintaining comparable power. For example, on the CHAMP dataset with $n_{a}$ of 1130, increasing $T$ from 100 to 500 samples reduced $\alpha$ from 6.3\% to 5.8\% while preserving power, and on the MGTX dataset with $n_{a}$ of 40, increasing $T$ from 100 to 1K samples reduced $\alpha$ from 20.7\% to 5.2\% while power decreased from 89.7\% to 87.4\%. Therefore, since the significance levels of \emph{Standard} are actually near 5\% (based on more trials) and those of SECRETS converge to 5\% while maintaining power (by increasing $T$), SECRETS is expected to outperform \emph{Standard} in these cases, as predicted by the power estimation model as well (e.g., CHAMP with $n_{a}$ of 1130 in Table \ref{tab:champ}). 

\subsection*{Ablation Studies on SECRETS}

Having demonstrated the ability of SECRETS to increase power and thereby reduce sampling complexity, we assess the 
importance of each design choice in the framework. First, we show that ITE estimation with SI outperforms ITE estimation 
with VT, by comparing the performance of SECRETS against SECRETS-VT. Then, we show that our data-driven hypothesis 
testing strategy is essential to the performance of SECRETS by comparing against alternative standard test strategies. 
We also show that our test strategy performs well, converging close to the performance obtained by an oracle version 
of the strategy.

\subsubsection*{Synthetic Intervention vs. Virtual Twins}

Compared to SECRETS, SECRETS-VT has higher variance in the ATE estimate, giving it 
lower power but comparable significance levels, in general. For example, on the CHAMP dataset (Table \ref{tab:champ}), for an arm size of
550, SECRETS and SECRETS-VT both have significance levels near 5\% but score 78\% and 58\% power, respectively. Applying the power equation model shows that, despite better separating the means, SECRETS-VT increases the variance of the ATE estimates, thereby lowering its \emph{shift term} compared to that under SECRETS. Similar trends hold on the ICARE dataset (Table \ref{tab:icare}). Although the model is unsuitable for the MGTX dataset given its small sample size, SECRETS-VT still has higher variance than SECRETS, likely explaining the reduced power (Table \ref{tab:mgtx}).

\subsubsection*{Data-driven Hypothesis Testing vs. Alternative Standard Hypothesis Testing Strategies} \label{sec:hyp_testing_ablation}

\begin{table*}[h]
\centering
\caption{Significance levels, $\alpha$ (\%) under different hypothesis testing strategies}
\label{tab:hyp_test} 
\resizebox{\columnwidth}{!}{%
\begin{tabular}{|M{0.1\linewidth} M{0.1\linewidth} M{0.175\linewidth} M{0.175\linewidth} M{0.225\linewidth} M{0.175\linewidth} M{0.175\linewidth}|}
\hline
Dataset & $n_{a}$ & SECRETS-T & SECRETS-B & SECRETS-T-P & SECRETS & SECRETS-O \\
\hline
\multirow[c]{2}{*}{CHAMP}
& 550 & 2.2 & 2.5 & 6.8 & 5.1 & 5.0 \\
& 1130 & 5.3 & 5.6 & 5.4 & 6.3 & 5.0 \\ \hline
\multirow[c]{2}{*}{ICARE}
& 1250 & 0.0 & 0.0 & 4.3 & 5.5 & 5.0 \\
& 5207 & 3.5 & 3.5 & 5.8 & 5.7 & 5.0 \\ \hline
\multirow[c]{2}{*}{MGTX}
& 30 & 0.0 & 0.0 & 5.3 & 5.7 & 5.1 \\
& 40 & 0.0 & 0.0 & 5.3 & 5.2 & 5.1 \\ \hline
{Average} & - & 1.8 & 1.9 & 5.5 & 5.6 & 5.0 \\ \hline 
\end{tabular}%
}
\end{table*}

Next, we show that our data-driven hypothesis test, which accounts for dependencies among the samples (i.e., the estimated ITEs), is necessary for achieving the desired significance level $\alpha_{target}$ (set to 5\%) 
in order to maximize power. To do this, we compare the significance levels under SECRETS against those under the 
standard hypothesis testing strategies, i.e., SECRETS-T, SECRETS-B, and SECRETS-T-P, with results reported in 
Table \ref{tab:hyp_test}. SECRETS-T, which uses the one-sample \emph{t}-test, achieves an average $\alpha$ of 
1.8\% across the datasets and arm sizes ($n_{a}$), which suggests that the test's assumptions, i.e., observations 
are drawn independently and identically from a Gaussian distribution, fail to hold. First, we check if Gaussianity 
is violated by swapping the one-sample \emph{t}-test with the bootstrap hypothesis test, i.e., SECRETS-B. SECRETS-B 
obtains a slightly higher average $\alpha$ of 1.9\%, implying that non-Gaussianity is not the problem. 
To show that the samples are dependent, we run a version of the one-sample \emph{t}-test, 
i.e., SECRETS-T-P, in which the estimated ITEs are shuffled across the trials to remove the sample dependencies 
existing within a single trial. SECRETS-T-P obtains an average $\alpha$ of 5.5\%, which is significantly closer to 
$\alpha_{target}$; this result affirms that the sample dependency induced under SI is the  violation incurred 
under standard hypothesis tests. To address the sample dependency problem, SECRETS implements a hypothesis test that uses 
the data to construct a null distribution that captures the dependencies in the data and thereby obtains an average $\alpha$ of 5.6\%. Increasing $T$ is expected to help 
significance levels of SECRETS converge to 5\% by fitting more accurate null distributions. For example, on the CHAMP dataset with $n_{a}$ of 1130, increasing $T$ from 100 to 500 samples reduced $\alpha$ from 6.3\% to 5.8\%, and on the MGTX dataset with $n_{a}$ of 40, increasing $T$ from 100 to 1K samples reduced $\alpha$ from 20.7\% to 5.2\%. In addition, the oracle version of SECRETS, i.e., SECRETS-O, which constructs the null 
distribution using ITEs across \emph{all} the trials, obtains nearly 5\% significance level across all datasets and arm sizes, 
suggesting that SECRETS can benefit from more diverse samples of ITEs under the null hypothesis, which may be obtained 
by increasing $T$.

\section{Discussion and Future Work} \label{sec:discuss}

Given the results of our experiments, we first discuss their implications and then suggest avenues for future work. 

\subsection*{Why is SECRETS effective?}
Our experiments show that SECRETS is able to improve the power while maintaining comparable significance level at a given sample size over the baseline 
approach by consistently reducing the variance of ATE estimates. It achieves this by simulating the cross-over design; specifically, 
it \emph{estimates} the ITEs across all units over the control and treatment groups using SI and averages over the 
estimated ITEs. Our ablation studies on the estimator attest that SI is better able to reduce the variance of the 
distributions compared to standard regression (i.e., the VT algorithm). SI's advantage likely stems from its 
nonparametric framing of the problem, in which it learns how to weight responses of donor units to predict the target unit's
response, unlike the VT algorithm that predicts responses using parametric models based only on the baseline measurement. While 
SI is important for reducing the variance of the ATE estimates, SECRETS also needs to 
use a hypothesis testing algorithm applicable to the estimated ITEs, in order to achieve the desired significance level 
and maximize power. Since SI predicts counterfactuals for a target unit using all units from the desired treatment arm, 
the estimated ITEs for target units from the same arm become dependent, hence rendering standard hypothesis tests 
unsuitable, as shown through our hypothesis testing ablation studies. To overcome the violations against sample 
independence under standard hypothesis tests, SECRETS uses 
a novel data-driven hypothesis testing algorithm, which constructs a null distribution that models the dependencies 
within the estimated ITEs, without making any parametric assumptions on their underlying distribution. Hence, the 
combination of an effective counterfactual estimator and a suitable hypothesis testing scheme is what ``powers" SECRETS.         
\subsection*{Future Work}

To increase the efficacy and utility of SECRETS, we suggest some directions for further research. First, the 
counterfactual estimation algorithm can be improved to reduce the errors in the ATE estimates, especially under the 
alternative setting, since the ATE estimate may also be of interest in the study. 
One approach towards this may be to learn an estimator based on all the measured baseline covariates, given that 
clinical RCTs typically collect numerous ordinal and numerical measurements on categories like demographics, medical 
history, and clinical assessments (e.g., the RCTs we examined had 100s-1000s of such covariates). While SI can 
incorporate multiple covariates by simply augmenting the pre-intervention data with the additional covariates, this strategy needs to be adapted to 
learn across different covariate types, i.e., numerical and ordinal covariates, since the regression setup differs 
across the covariate types \cite{harrell2001regression}. In addition to improving the estimator's performance, 
SECRETS can be made more useful for study planning by extending it with a module that estimates the sample size 
required for a desired significance level and power.  

\section{Conclusion} \label{sec:con}

In conclusion, we have developed SECRETS, a novel framework that, for the first time, increases the power and
consequently decreases the sampling complexity of a clinical RCT, \emph{without relying on any external data}, as
done in prior approaches. 
SECRETS achieves this by
simulating the cross-over design; it estimates ITEs per patient across \emph{both} control and treatment groups in
order to reduce the variance of the ATE estimate, whereas prior approaches only do so for the treatment group.
To estimate ITEs for patients across both groups, it uses SI, a
state-of-the-art counterfactual estimation algorithm. Then, to leverage the reduced variance of the ATE estimates
in order to increase power, it implements a novel data-driven hypothesis testing strategy suitable for the
estimated ITEs since their properties violate the independence assumption under standard hypothesis test schemes.
Evaluated on three real-world Phase-3 clinical RCTs, i.e., the CHAMP, ICARE, and MGTX studies, SECRETS improves 
power over the baseline approach by 6-54\% with an average improvement of 21.5\% (standard deviation of 15.8\%) while maintaining comparable significance levels. 
In addition, the gains in power reduce the number of subjects required to obtain a typically desired statistical 
operating point of 80\% power and 5\% significance level by 
51\% or 580 subjects per arm on the CHAMP dataset, 76\% or 3957 subjects per arm on the ICARE dataset, and 25\% or 10 subjects per arm on the MGTX dataset. While 
SECRETS can currently be used to boost the power of an already conducted clinical RCT, our future work will look 
into improving the counterfactual estimation algorithm to reduce errors in ATE estimates and developing a method 
for sample size estimation under SECRETS as a study planning aide.  


\section*{Acknowledgment}

This work was supported by the NSF under Grant No. CNS-1907381 and was performed using resources from Princeton Research Computing. 

This research is based on data from NINDS obtained from its Archived Clinical Research Dataset website. The CHAMP dataset was obtained from the Childhood and 
Adolescent Migraine Prevention Study, conducted under principal investigators (PIs) Drs. Powers, Hershey, and Coffey, under Grant 
\#1U01NS076788-01. The ICARE dataset was obtained from the Arm Rehabilitation Study After Stroke (ICARE), conducted 
under PIs Drs. Winstein, Dromerick, and Wolf, under Grant \#U01NS056256. The MGTX dataset was obtained from the Thymectomy Trial in 
Non-Thymomatous Myasthenia Gravis Patients Receiving Prednisone Therapy, conducted under 
PIs Drs. Cutter, Wolfe, and Kaminski under Grant \#1U01NS042685-01A2.

The authors thank Dr. Wolfe for his guidance on RCTs and the MGTX study.


%
%

 \vspace*{ 1 cm}

\bibliographystyle{IEEEtran}
\bibliography{references}

\clearpage

\appendix

\section{SECRETS Pseudocode} \label{sec:supp-alg}

 \begin{algorithm*}[hbt!] 
     \caption{\textproc{SECRETS}}
     \label{alg:main}
     \begin{algorithmic}[1]
     \Statex \textbf{Input}:
     \Statex $X_{ctrl} \in \mathbb{R}^{n_{a}\times n_{t}}$: control group data, where $n_{a}$ is the arm size, $n_{t}$ is the duration of the study, where $t=1$ is the pre-intervention timepoint and $t>1$ is the post-intervention period 
     \Statex $X_{treat} \in \mathbb{R}^{n_{a}\times n_{t}}$: treatment group data, where $n_{a}$ is the arm size, $n_{t}$ is the duration of the study, where $t=1$ is the pre-intervention timepoint and $t>1$ is the post-intervention period
     \Statex  SI\_tuning\_params: dictionary containing the following arguments for SI tuning
     \Statex\hspace{\algorithmicindent} $r_{train,val}$: ratio of training to validation set size for tuning SI's hyperparameters
     \Statex $get\_outcome: \mathbb{R}^{n_{t}} \rightarrow \mathbb{R}$: function that calculates the outcome of interest from a patient's response trajectory under some intervention over the study duration $n_{t}$ 
     \Statex $T$: number of samples to generate from the null distribution
	\Statex  Testing\_params: dictionary containing the following arguments for tuning the test's critical value
     \Statex\hspace{\algorithmicindent} $\alpha_{target}$: target significance level
     \Statex\hspace{\algorithmicindent} $t_{lower}$: critical value search lowerbound; $\geq 0$ (because of two-sided testing) 
     \Statex\hspace{\algorithmicindent} $t_{upper}$: critical value search upperbound; $\geq t_{lower}$
     \Statex\hspace{\algorithmicindent} $t_{limit,exp}$: critical value search limit expansion term
     \Statex\hspace{\algorithmicindent} $n_{s}$: number of candidate critical values to search over
     \Statex\hspace{\algorithmicindent} $\delta_{\alpha_{target}}$: significance level error tolerance
     \Statex \textbf{Output}:
     \Statex $d$: test outcome, 1 means reject and 0 means do not reject the null hypothesis 
     \State $Y_{ctrl}=\Call{estimate\_ites}{X_{ctrl}, X_{treat},\text{SI\_tuning\_params},get\_outcome}$ \label{main:line:estimate_ites}
     \State $Y_{treat}=\Call{estimate\_ites}{X_{treat}, X_{ctrl},\text{SI\_tuning\_params},get\_outcome}$ \label{main:line:estimate_ites}
     \State $Y=concatenate(Y_{ctrl},Y_{treat})$
	 \State $d=\Call{run\_hypothesis\_test}{X_{ctrl},Y,\text{SI\_tuning\_params},get\_outcome,T,\text{Testing\_params}}$ \label{main:line:hyp_test}
     \State \textbf{return} $d$ \label{main:line:end}
     \end{algorithmic}
 \end{algorithm*}

\clearpage 

\begin{algorithm*}[hbt!]
\caption{\textproc{tune\_SI\_hyperparams}}
\label{alg:tune_si_hparams}
\begin{algorithmic}[1]
\Statex \textbf{Input}:
\Statex $X \in \mathbb{R}^{n_{a}\times n_{t}}$: data from a group exposed to a single intervention, where $n_{a}$ is the number of subjects in the group, $n_{t}$ is the duration of the study, where $t=1$ is the pre-intervention timepoint and $t>1$ is the post-intervention period
\Statex  params: dictionary containing the following arguments for SI tuning
\Statex\hspace{\algorithmicindent} $r_{train,val}$: ratio of the training to validation set size for tuning SI's hyperparameters
\Statex \textbf{Output}:
\Statex $\lambda_{ridge,best}$: tuned ridge regularization strength
\Statex $\lambda_{svt,best}$: tuned truncation threshold for singular value thresholding 

\State $X_{train},X_{val}=split\_train\_val(X,\text{params}.r_{train,val})$
\State $size,n_{t}=shape(X_{val})$

\State $best\_score=-\infty$ \label{est_ites_helper:line:init_params_s} 
\State $\lambda_{ridge,best}=\textsc{NIL}$
\State $\lambda_{svt,best}=\textsc{NIL}$ \label{est_ites_helper:line:init_params_e}

\For{$\lambda_{ridge} \in logspace(-3,3,7)$} \label{est_ites_helper:line:sweep_s} 
	\For{$\lambda_{svt} \in linspace(0.1,1,10)$}  
     
        \State $X_{val,pred}=\textbf{0}^{size \times n_{t}}$  \label{est_ites_helper:line:val_init_s}
        \For{$i=1,size$} \label{est_ites_helper:line:val_pred_s}
            \State $X_{val,pred}[i]$=\Call{SI}{$\lambda_{ridge}$,$\lambda_{svt}$,$X_{train}$,$X_{val}[i]$} 
        \EndFor \label{est_ites_helper:line:val_pred_e}

        \State $score=get\_rsquared(X_{val},X_{val,pred})$ \label{est_ites_helper:line:param_update_s}
        \If{$score > best\_score$} 
            \State $best\_score=score$
            \State $\lambda_{ridge,best}=\lambda_{ridge}$
            \State $\lambda_{svt,best}=\lambda_{svt}$
        \EndIf \label{est_ites_helper:line:param_update_e}
    \EndFor
\EndFor \label{est_ites_helper:line:sweep_e}

\State \textbf{return} $\lambda_{ridge,best},\lambda_{svt,best}$

\end{algorithmic}
\end{algorithm*}

\textproc{tune\_SI\_hyperparams} (Alg. \ref{alg:tune_si_hparams}) first initializes the set of best hyperparameters (lines 
\ref{est_ites_helper:line:init_params_s}-\ref{est_ites_helper:line:init_params_e}) and then sweeps over a wide range 
of candidate values for $\lambda_{ridge}$ and $\lambda_{svt}$, tracking the validation performance under each hyperparameter 
configuration (lines \ref{est_ites_helper:line:sweep_s}-\ref{est_ites_helper:line:sweep_e}), where the validation 
performance is the $R^{2}$ score between the true validation data and corresponding SI-derived 
predictions. To obtain the SI-derived predictions per unit, \textproc{tune\_SI\_hyperparams} runs \textproc{SI} (Alg. \ref{alg:si}) with the 
given candidate hyperparameters and $X_{train}$ as the donor data, and $X_{val}[i]$ as the target unit data 
(lines \ref{est_ites_helper:line:val_pred_s}-\ref{est_ites_helper:line:val_pred_e}). After estimating 
the trajectories across the validation units, \textproc{tune\_SI\_hyperparams} calculates the validation 
performance; if the validation performance improves, it updates the best hyperparameter configuration (lines \ref{est_ites_helper:line:param_update_s}-\ref{est_ites_helper:line:param_update_e}). It concludes by returning the best hyperparameter configuration over the search range.
 
\clearpage 

\begin{algorithm*}[hbt!]
\caption{\textproc{SI}}
\label{alg:si}
\begin{algorithmic}[1]

\Statex \textbf{Input}:
\Statex $\lambda_{ridge}$: ridge regularization strength
\Statex $\lambda_{svt}$: truncation threshold for singular value thresholding
\Statex $X_{donor} \in \mathbb{R}^{n_{d}\times n_{t}}$: donor data, where $n_{d}$ is the number of donor subjects, $n_{t}$ is the duration of the study, where $t=1$ is the pre-intervention timepoint and $t>1$ is the post-intervention period
\Statex $X_{unit} \in \mathbb{R}^{n_{t}}$: target unit data, where $n_{t}$ is the duration of the study, where $t=1$ is the pre-intervention timepoint and $t>1$ is the post-intervention period
\Statex \textbf{Output}:
\Statex $X_{unit,c} \in \mathbb{R}^{n_{t}}$: target unit's counterfactual data (under the intervention assigned to the donor group) over the study duration 

\State $n_{d},n_{t}=shape(X_{donor})$ \label{est_ites_helper:line:reshape_s}
\State $X_{unit}=reshape(X_{unit},(1,n_{t}))$ \label{est_ites_helper:line:reshape_e}

\State $U,s,V^{T}=\Call{SVD}{X_{donor}}$  \label{est_ites_helper:line:svd_s}
\State $S=\{i: s[i] \geq \lambda_{svt}\}$
\State $X_{donor,trunc}=\sum_{i\in S} s[i]U[i]V[i]^{T}$ \label{est_ites_helper:line:svd_e}
 
\State 
    $w = \argmin_w \lVert X_{unit}[:,1] - w^{T}X_{donor,trunc}[:,1] \rVert^{2}_{2} + \lambda_{ridge}\lVert w \rVert^{2}_{2} $ \label{est_ites_helper:line:reg_s} \label{est_ites_helper:line:reg_e}
\State $X_{unit,c}=\textbf{0}^{n_{t}}$
\State $X_{unit,c}[1]=squeeze(X_{unit})[1]$ \label{est_ites_helper:line:init_pre}
\State $X_{unit,c}[2:]=squeeze(w^{T}X_{donor,trunc}[:,2:])$ \label{est_ites_helper:line:pred_s}
\State \textbf{return} $X_{unit,c}$ \label{est_ites_helper:line:pred_e}

\end{algorithmic}
\end{algorithm*}

\textproc{SI} (Alg. \ref{alg:si})
estimates the counterfactual post-intervention trajectory for the target unit by denoising the donor data with singular value thresholding (lines 
\ref{est_ites_helper:line:svd_s}-\ref{est_ites_helper:line:svd_e}), learning the donor weights with ridge regression 
on the pre-intervention (i.e., baseline) data (line \ref{est_ites_helper:line:reg_s}), 
and using the learned donor weights to estimate the target unit's post-intervention trajectory from the donors' 
post-intervention data (line \ref{est_ites_helper:line:pred_s}). Since SI is used to predict the post-intervention trajectory of the target unit, the unit's ``counterfactual" pre-intervention data is initialized to its original pre-intervention data (line \ref{est_ites_helper:line:init_pre}).

\clearpage

\begin{algorithm*}[hbt!]
\caption{\textproc{tune\_critical\_value}}
\label{alg:get_cv}
\begin{algorithmic}[1]
\Statex \textbf{Input}:
\Statex $s_{null} \in \mathbb{R}^{T}$: $T$ samples from the null distribution of the test statistic 
\Statex params: dictionary containing the following arguments for tuning the test's critical value
     \Statex\hspace{\algorithmicindent} $\alpha_{target}$: target significance level
     \Statex\hspace{\algorithmicindent} $t_{lower}$: critical value search lowerbound; $\geq 0$ (because of two-sided testing) 
     \Statex\hspace{\algorithmicindent} $t_{upper}$: critical value search upperbound; $\geq t_{lower}$
     \Statex\hspace{\algorithmicindent} $t_{limit,exp}$: critical value search limit expansion term
     \Statex\hspace{\algorithmicindent} $n_{s}$: number of candidate critical values to search over
     \Statex\hspace{\algorithmicindent} $\delta_{\alpha_{target}}$: significance level error tolerance
    \Statex \textbf{Output}:
    \Statex $d$: test outcome, 1 means reject and 0 means do not reject the null hypothesis

\Procedure{Get\_alpha}{$s_{null}, t$} \label{get_cv:line:get_alpha_s}
    \State $\alpha=0$ \label{get_cv:line:init_alpha_s}
    \For{$i=1,length(s_{null})$} \label{get_cv:line:est_alpha_s}
        \State $r=\Big|s_{null}[i]\Big|>t$ \label{get_cv:line:test_s}
        \State $\alpha=\alpha+r$
    \EndFor \label{get_cv:line:est_alpha_e}
    \State \textbf{return} $\alpha/length(s_{null})$ \label{get_cv:line:frac}
    \EndProcedure \label{get_cv:line:get_alpha_e}

    \State $t_{candidates}=linspace(\text{params.}t_{lower},\text{params.}t_{upper},\text{params.}n_{s})$ \label{get_cv:line:search_init_s}
    \State $\hat{\alpha}\_by\_t=[]$ \label{get_cv:line:search_init_e}
    \For{$i=1, \text{params.}n_{s}$} \label{get_cv:line:search_s}
        \State $t=t_{candidates}[i]$ \label{get_cv:line:eval_t_s}
        \State $\hat{\alpha}$=\Call{Get\_alpha}{$s_{null},t$} \label{get_cv:line:get_alpha}

        \If{$|\hat{\alpha}-\text{params.}\alpha_{target}|<\text{params.}\delta_{\alpha_{target}} $} \label{get_cv:line:stop_s}
            \State \textbf{return} $t$
        \EndIf \label{get_cv:line:stop_e}

        \State $\hat{\alpha}\_by\_t[i]=\hat{\alpha}$ \label{get_cv:line:eval_t_e} 
    \EndFor \label{get_cv:line:search_e}


    \State $m=\text{params.}n_{s}/2$ \label{get_cv:line:midpt_s}
    \State $\alpha_m=\hat{\alpha}\_by\_t[m]$ \label{get_cv:line:midpt_e}
    \If{$\text{params.}\alpha_{target} < \alpha_m$} \label{get_cv:line:alpha_lower_s}
        \State $\text{params.}t_{lower}=t_{candidates}[m+1]$ \label{get_cv:line:alpha_lower_e}
        \If{$\text{params.}\alpha_{target} < \hat{\alpha}\_by\_t[-1]$} \label{get_cv:line:alpha_lower_tu_s}
            \State $\text{params.}t_{upper}=\text{params.}t_{upper}+\text{params.}t_{limit,exp}$  \label{get_cv:line:alpha_lower_tu_e}
        \EndIf 
    \Else \label{get_cv:line:alpha_higher_s}
        \State $\text{params.}t_{upper}=t_{candidates}[m+1]$ \label{get_cv:line:alpha_higher_e}
        \If{$\text{params.}\alpha_{target} > \hat{\alpha}\_by\_t[0]$} \label{get_cv:line:alpha_higher_tl_s}
            \State $\text{params.}t_{lower}=max(0,\text{params.}t_{lower}-\text{params.}t_{limit,exp})$  \label{get_cv:line:alpha_higher_tl_e}
        \EndIf 
    \EndIf 
    
    \State \textbf{return} \Call{tune\_critical\_value}{$s_{null}$, $params$} \label{get_cv:line:binary_search_e}
\end{algorithmic}
\end{algorithm*}

\clearpage 

\textproc{tune\_critical\_value} (Alg.~\ref{alg:get_cv}) searches over critical values to find the one yielding significance level $\text{params.}\alpha_{target}$. First, it sweeps over a range of candidate critical values $t_{candidates}$, defined by $\text{params.}n_{s}$, $\text{params.}t_{lower}$, 
and $\text{params.}t_{upper}$, and checks if any 
value obtains the desired significance level $\text{params.}\alpha_{target}$ (lines \ref{get_cv:line:search_init_s}-\ref{get_cv:line:search_e}). For 
each $t$, it estimates the resulting significance level $\hat{\alpha}$ by calling \textproc{Get\_alpha} (line \ref{get_cv:line:get_alpha}), which calculates $\hat{\alpha}$ by evaluating the 
two-sided test with the candidate critical value $t$. \textproc{Get\_alpha} (lines \ref{get_cv:line:get_alpha_s}-\ref{get_cv:line:get_alpha_e}) runs the test on each sample from the null distribution $s_{null}$ (line \ref{get_cv:line:test_s}) and calculates 
the fraction of samples on which the test returns a reject (line \ref{get_cv:line:frac}). If the critical value yields $\hat{\alpha}$ 
close to $\text{params.}\alpha_{target}$ (i.e., based on the error tolerance $\text{params.}\delta_{\alpha_{target}}$), 
\textproc{tune\_critical\_value} returns the critical value (lines \ref{get_cv:line:stop_s}-\ref{get_cv:line:stop_e}); otherwise, it stores the corresponding significance level and resumes 
the sweep (line \ref{get_cv:line:eval_t_e}). 

If none of the candidate values yield 
significance levels close to $\text{params.}\alpha_{target}$, \textproc{tune\_critical\_value} updates its search range in a fashion similar to 
binary search. First, it identifies the significance level of the middle candidate critical value $\alpha_{m}$ 
(lines \ref{get_cv:line:midpt_s}-\ref{get_cv:line:midpt_e}). If $\text{params.}\alpha_{target}$ is less than $\alpha_{m}$, 
\textproc{tune\_critical\_value} updates its search range to the right half of the original critical value search space since 
increasing the critical value would decrease the significance level 
(lines \ref{get_cv:line:alpha_lower_s}-\ref{get_cv:line:alpha_lower_e}). In addition, if $\text{params.}\alpha_{target}$ is less 
than the $\hat{\alpha}$ associated with $\text{params.}t_{upper}$, $\text{params.}t_{upper}$ needs to be sufficiently increased (i.e., by 
$\text{params.}t_{limit,exp}$) to ensure the updated search range contains a critical value yielding $\alpha_{target}$ (lines 
\ref{get_cv:line:alpha_lower_tu_s}-\ref{get_cv:line:alpha_lower_tu_e}). Analogously, if $\text{params.}\alpha_{target}$ is greater 
than $\alpha_{m}$, \textproc{tune\_critical\_value} updates its search range to the left half of the original critical value search 
space since decreasing the critical value would increase the significance level (lines 
\ref{get_cv:line:alpha_higher_s}-\ref{get_cv:line:alpha_higher_e}). In addition, if $\text{params.}\alpha_{target}$ is greater 
than $\hat{\alpha}$ associated with $\text{params.}t_{lower}$, $\text{params.}t_{lower}$ needs to be sufficiently decreased (i.e., by 
$\text{params.}t_{limit,exp}$ but lowerbounded by 0 because of two-sided testing) to ensure the updated search range contains 
a critical value yielding $\alpha_{target}$ (lines \ref{get_cv:line:alpha_higher_tl_s}-\ref{get_cv:line:alpha_higher_tl_e}). After updating the 
search range, \textproc{tune\_critical\_value} continues searching by recursing with the updated search parameters and returning 
the identified critical value (line \ref{get_cv:line:binary_search_e}).

\clearpage

\section{Clinical RCT Datasets} \label{sec:supp-datasets}

\subsection{CHAMP}

The CHAMP study \cite{powers2017trial} evaluated whether different medications could reduce headache frequency and heachache effects among 
children and adolescents suffering from migraines. To assess this, the study implemented an RCT containing a placebo 
group and two treatment groups receiving amitriptyline and topiramate, respectively. At the end of the trial, 
treatment effects were measured on various outcomes, including change in headache frequency, number of headache days, 
and headache-related disability scores, all relative to baseline measurements. Based on data from 328 subjects 
collected over 24 weeks, the study did not find any clinically significant between-group differences across the 
health outcomes. 

For our experiments, we use the change in the Pediatric Migraine Disability Assessment (PedMIDAS) score as the outcome metric 
defining the ATE and the amitriptyline and topiramate as the control and treatment groups, respectively, because the 
corresponding ATE was more statistically significant among the ATEs defined by other continuous metrics and arm 
pairs \cite{powers2017trial,champ_stats}. To calculate this ATE, the study analyzed a subset of subjects monitored over two
visits, i.e., one at baseline and another near the 24-week endpoint, which comprised 211 subjects with 107 and 104 subjects in the amitriptyline and topiramate groups, respectively. After applying the same criterion, we 
extracted 204 subjects from the original dataset, with 106 in the amitriptyline group and 98 in the topiramate group. From our extracted dataset, we calculated the 
ATE to be -3.17 units, somewhat comparable to the -4.3 units reported in the study \cite{champ_stats}.  

\subsection{ICARE}

The ICARE study \cite{winstein2016effect} evaluated whether a new motor training program (Accelerated Skill Acquisition Program or ASAP) could reduce upper extremity disability among 
patients with motor stroke more effectively than usual customary care (UCC). To assess this, the study implemented an 
RCT containing a treatment group exposed to ASAP, a control group exposed to dose-equivalent usual customary care 
(DEUCC), and another control group exposed to UCC with no constraint on the dose. At the end of the trial, treatment 
effects were assessed over various outcomes, including changes in Wolf Motor Function Test (WMFT) time, Stroke Impact 
Scale (SIS) scores, and arm muscle torque. Based on data from 304 patients collected over one year, the study did not 
find any clinically significant between-group differences in a subset of these scores, i.e., WMFT and SIS hand 
function score \cite{winstein2016effect, icare_stats}. 

For our experiments, we use the change in arm muscle torque based on shoulder flexors as the outcome metric defining the ATE, 
and the DEUCC and ASAP as the control and treatment groups, respectively, to speed up experiment time since detecting 
this ATE with high power and low significance level required a relatively small sample size for the baseline method.  
After applying the study's data processing protocol to the original dataset, we extracted data from 183 subjects, with 93 in the control group and 90 in the treatment 
group, from visits at the baseline and one-year endpoints (the number of subjects analyzed per group is equal to that reported in the study's analysis). From our extracted dataset, we calculated the ATE to 
be -3.00 units, close to the -2.99 reported in the study \cite{icare_stats}.

\subsection{MGTX}

The MGTX study \cite{wolfe2016randomized} investigated whether thymectomy combined with standard prednisone therapy could treat Myasthenia Gravis 
more effectively than prednisone therapy alone. To assess this, the study implemented an RCT, in which the control 
arm received prednisone therapy over three years and the treatment arm underwent thymectomy and received the same 
prednisone therapy; at the end of the trial, treatment effects were measured with respect to the Quantitative 
Myasthenia Gravis (QMG) total score and required prednisone dose, both averaged over the study period. Based on data 
from 126 subjects collected over three years, the study found that thymectomy improved health outcomes with clinical 
significance; the time-weighted average QMG scores decreased by an average of 2.85 units and the time-weighted average 
prednisone dose also decreased by an average of 22 mg. 

For our experiments, we use the time-weighted average QMG total score as the outcome metric defining the ATE, because it was 
easier to reproduce \cite{wolfe2016randomized}. The study calculated the ATE by analyzing a subset of the population monitored over the three-year window (14 patient visits or timepoints), which comprised 62 subjects and 56 subjects in the treatment and control groups, respectively. We followed the study's data processing protocol and from the 
original dataset, we extracted a dataset with 49 subjects in the treatment group and 47 subjects in 
the control group. From our extracted dataset, 
we calculated the ATE to be -2.70 units, comparable to the -2.85 units reported in the study.

\section{Algorithmic Parameters} \label{sec:supp-alg_params}

We define the \emph{get\_outcome} function, used by \emph{Standard} and SECRETS, for each RCT dataset
according to the corresponding outcome metric specified in Appendix \ref{sec:supp-datasets}. For \emph{Standard}, we set the target significance level $\alpha_{target}=5\%$. For SECRETS, we set the parameters contained in SI\_tuning\_params as follows: $r_{train,val}$ to 7/3. We set $T$ to 100 for CHAMP and ICARE and 1000 for MGTX. We set the parameters contained in Testing\_params as follows: $\alpha_{target}=5\%$, $t_{lower}=3$, $t_{upper}=5$, $t_{limit,exp}=2$, $n_{s}=10$, and $\delta_{\alpha_{target}}=$ $1e$-$3$. We found this parameter configuration enabled SECRETS to achieve significance level close to $\alpha_{target}$. See Appendix \ref{sec:supp-alg}, Alg. \ref{alg:main} for descriptions of these parameters.

\section{Tables} \label{sec:supp-tables}

\setcounter{table}{0}

\begin{table*}[h]
\centering
\caption{Effect of Number of Trials ($L$) on \emph{Standard}'s performance}
\label{tab:num_trials} 
\resizebox{\columnwidth}{!}{%
\begin{tabular}{|M{0.22\linewidth} M{0.2\linewidth} M{0.2\linewidth} M{0.2\linewidth} M{0.2\linewidth}|}
\hline

\multirow[c]{2}{*}{Dataset} 
	& \multicolumn{2}{c}{$L=$1K} & \multicolumn{2}{c|}{$L=$10K} \\
	& $1-\beta$ (\%) & $\alpha$ (\%) & $1-\beta$ (\%) & $\alpha$ (\%) \\ \hline 

					

CHAMP, $n_{a}=1130$  & 81 & 3.8 & 81 & 4.5 \\ \hline 
ICARE, $n_{a}=1250$ & 27 & 4.1 & 28 & 4.9  \\ \hline 
ICARE, $n_{a}=5207$ & 81 & 5.8 & 80 & 5.5 \\ \hline 
\end{tabular}%
}
\end{table*}

\end{document}